\documentclass[prl,aps,twocolumn,showpacs,superscriptaddress]{revtex4}
\usepackage{epsfig}

\begin{document}

\author{Eugeniy E.\ Mikhailov}
\affiliation{
	Department of Physics and Institute of Quantum Studies,
	Texas A\&M University,
	College Station, Texas 77843-4242
}
\author{Vladimir A.\ Sautenkov}
\affiliation{
	Department of Physics and Institute of Quantum Studies,
	Texas A\&M University,
	College Station, Texas 77843-4242
}
\affiliation{
	P.  N.  Lebedev Institute of Physics, 119991 Moscow, Russia
}
\author{Yuri V.\ Rostovtsev}
\affiliation{
	Department of Physics and Institute of Quantum Studies,
	Texas A\&M University,
	College Station, Texas 77843-4242
}
\author{Aihua Zhang}
\affiliation{
	Department of Physics and Institute of Quantum Studies,
	Texas A\&M University,
	College Station, Texas 77843-4242
}
\author{M.\ Suhail Zubairy}
\affiliation{
	Department of Physics and Institute of Quantum Studies,
	Texas A\&M University,
	College Station, Texas 77843-4242
}
\affiliation{
	Department of Electronics,
	Quaid-i-Azam University,
	Islamabad, Pakistan
}
\author{Marlan O.\ Scully}
\affiliation{
	Department of Physics and Institute of Quantum Studies,
	Texas A\&M University,
	College Station, Texas 77843-4242
}
\affiliation{
	Department of Chemistry,
	Princeton University,
	Princeton, NJ 08544
}
\author{George R.\ Welch}
\affiliation{
	Department of Physics and Institute of Quantum Studies,
	Texas A\&M University,
	College Station, Texas 77843-4242
}
\title{Spectral Narrowing via Quantum Coherence}

\begin{abstract}
	We have studied the transmission through an optically thick
	$^{87}$Rb vapor that is illuminated by monochromatic and noise broaden 
	laser fields in $\Lambda$ configuration.  The spectral width of the
	beat signal between the two fields after transmission
	through the atomic medium is more than 1000 times
	narrower than the spectral width of this signal before
	the medium.
\end{abstract}
\date{\today}
\maketitle

\newcommand{\ds}{\displaystyle}
\newcommand{\dd}{\partial}
\newcommand{\be}{\begin{equation}}
\newcommand{\ee}{\end{equation}}
\newcommand{\dt}{\ds\frac{\dd}{\dd t}}
\newcommand{\dz}{\ds\frac{\dd}{\dd z}}
\newcommand{\D}{\ds\left(\frac{\dd}{\dd t} + c \frac{\dd}{\dd z}\right)}

\newcommand{\w}{\omega}
\newcommand{\W}{\Omega}
\newcommand{\g}{\gamma}
\newcommand{\G}{\Gamma}
\newcommand{\E}{\hat E}
\newcommand{\s}{\sigma}


The resonant interaction of an quasi-monochromatic electromagnetic field 
with atomic media is very important because of 
its applications to spectroscopy, magnetometry~\cite{harris90prl,sz}, 
nonlinear optics~\cite{nlo}
and quantum information and computing~\cite{qc,qs}. 
%
%
The interaction of a phase noise broadened electromagnetic field 
with media possessing a resonant absorption or transmission 
has been studied both
experimentally and theoretically.  
In particular, several experiments involving
noisy laser fields transmitted through a cell containing
alkali atomic vapor were performed.  For example, the study
of the conversion of phase-noise to amplitude noise in dense
Cs~\cite{bahoura01} and Rb~\cite{zibrov_pros} confirmed the
theoretical prediction in~\cite{armstrong66}.  In addition,
there are experimental studies of intensity fluctuations and 
correlations between the drive and probe fields in the 
electromagnetically induced transparency (EIT)
regime~\cite{nusseinzveig02,lukin05,sautenkov05pra,huss04prl}.  


	In this letter, we show that the spectral width of the
beat signal between two lasers is modified by transmission
through a medium to give a very narrow spectral feature
that is more than 1000 times narrower than the spectral width of the probe
laser radiation.  
This can find broad applications to spectroscopy to develop 
light sources that have a very narrow (less than natural) spectral line  
ranging their carry frequency from optics to gamma-rays
where experimental demonstration of EIT has been recently 
reported \cite{g-eit}.
It is important to note here that the spectral width of the line 
and the coherent time of radiation is controlled by auxiliary 
external laser field. 
It is also worth to mention that the current technique can be applied to a 
single photon source (see for example \cite{lukin05}) 
and allows one to transfer the flux of
single photons with a particular or not defined 
coherent time to the flux of single photons
with a given coherent time.

\begin{figure*}
    \includegraphics[
	width=1.50\columnwidth
    ]
    {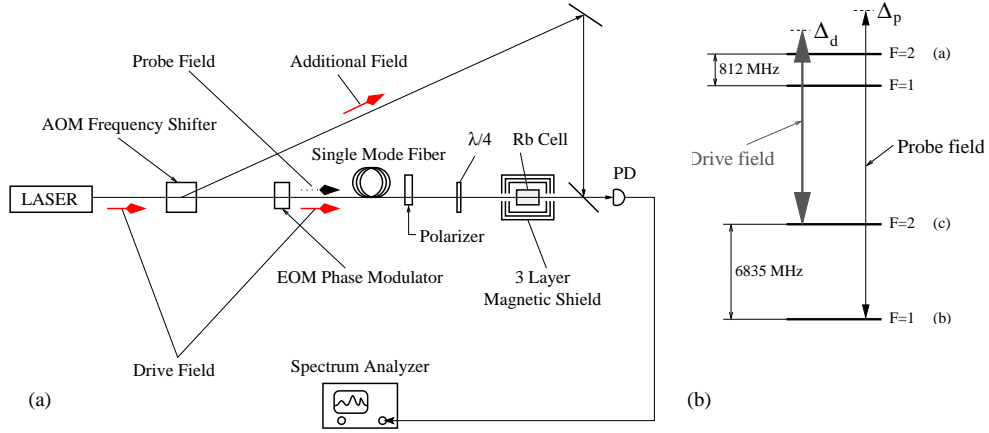}
    \caption{
	(a) Experimental setup.
        (b) Atomic level structure for $^{87}$Rb.
	\label{setup} \label{level-structure}
    }
\end{figure*}

Our experimental setup is shown in Fig.~\ref{setup}a.
The laser is tuned to the transition 
[(a) F=2 -- (c) F=2, see in Fig.~\ref{setup}b], 
and the output laser beam is modulated
by an electro-optic modulator (EOM) 
which is driven at the frequency of the ground
level splitting (6.835~GHz, see Fig.~\ref{setup}b).  
Two sidebands are generated,
one with the frequency of the probe field and another with
frequency down-shifted by 6.835 GHz with respect to the
drive field. This down-shifted field is far from resonance
and has a negligible effect on the experiment.  The power
and frequency of the sidebands is varied by changing the
frequency and amplitude of the microwave field driving the EOM.
If the EOM is driven by a spectrally narrow microwave source,
then we observe a narrow EIT spectrum, just as for the case
when two phase-locked lasers are used~\cite{kash99prl}.

After the EOM, optical fields are sent through a single mode 
optical fiber to make a clean spatial intensity distribution
with diameter 0.7 cm. The optical fields are circularly polarized by
a quarter-wave plate. Then the drive and phase-noise broadened probe beams 
propagate through a cell containing atomic $^{87}$Rb at temperature
$67.7C$ and a buffer gas neon at pressure 30 Torr; 
the length of the cell is 2.5 cm; the diameter is 2.2 cm.  
The power at the entrance of the cell is 0.58 mW; 
after the cell power is 0.32 mW. 
We use heterodyne detection of the probe by mixing the transmitted
light with an additional field that is frequency shifted by
an acousto-optic modulator (AOM) 
at 60~MHz with respect to the drive field.  This field
does not propagate through the $^{87}$Rb medium.  
This detection technique has been described in \cite{kash99prl}. 
A spectrum analyzer tuned in the vicinity of the beat-note frequency 
of the probe and additional fields is used to record the spectrum of 
beat-note which coincides with the spectral density of 
the phase-noise broadened transmitted probe field for the case of 
monochromatic drive field.

	In the current experiment, the EOM is modulated with a
broad (``white'') noise spectrum centered about the selected 
probe frequency.  The total power in each side-band is about
10\% of the power in the drive field, and the spectral width of
the beat signal between the probe and drive is $\sim$ 1~MHz.
We can characterize the modulation of the EOM as a time
dependent frequency:
$
f(t)=f_0+\Delta f(t)
$
where $f_0$ is the carrier frequency (6.835~GHz), and $\Delta
f(t)$ is the noise driven time-dependent frequency shift.  Since
the phase of such an oscillation is equal to
$
\varphi(t) = \varphi_0 + f_0 t + \int_{-\infty}^{t} \Delta f(t) dt,
$
we have a phase-noise broadened microwave signal that drives the EOM.
Thus we have a phase-noise broadened probe field instead of a
monochromatic probe field which is observed without the phase
noise modulation.  This is equivalent to using two lasers,
one of which is spectrally narrow, and the other of which is
has a phase-noise broadened spectrum.

\begin{figure}
    \includegraphics[
	width=0.90\columnwidth
    ]
    {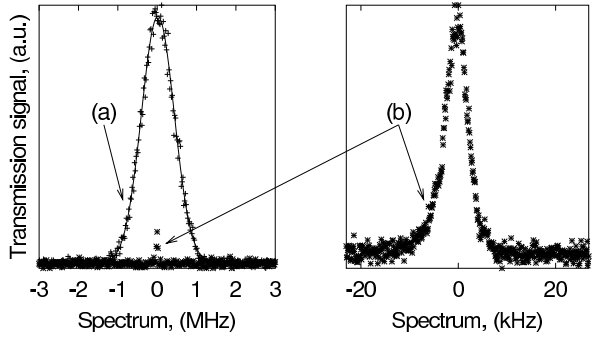}
    \caption{
%
%
%
	Spectral density of the phase-noise broadened probe
	laser after passing the $^{87}$Rb cell.  (a) shows the
	spectrum with the lasers far from resonance, so
	that the interaction with the atoms is negligible.
	(b) shows the spectrum with the lasers on resonance.
	The solid line in plot (a) is a Gaussian fit.
%
%
	\label{noise_in_out}
    }
\end{figure}
\begin{figure}
    \includegraphics[
	width=0.90\columnwidth
    ]
    {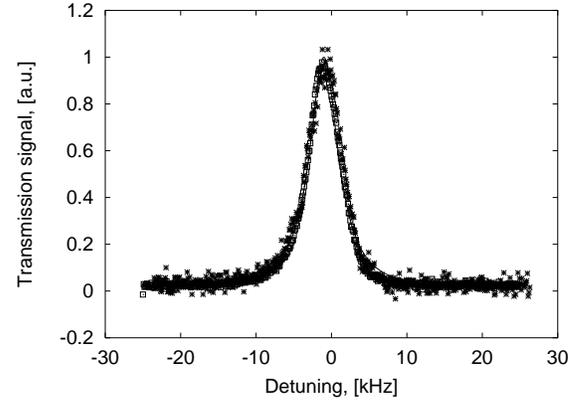}
    \caption{
	Comparison of the EIT resonance obtained two ways.
	Data points marked with squares are transmission versus
	probe laser frequency for spectrally narrow drive and
	probe fields.  Data points marked with stars are the
	spectrum of the transmitted probe light when the probe
	is phase-broadened.
	\label{noise_and_canonical_eit}
    }
\end{figure}

	We note that the spectral density of the probe
signal does not coincide exactly with the spectrum of driving microwave signal
because of the limited bandwidth of the EOM
response to the modulation frequency.  
A typical plot of spectral density is depicted in
Fig.~\ref{noise_in_out}.  Before the cell, the spectral
FWHM (full width at half maximum) of the probe field
is 980~kHz, whereas after the cell we see significant
narrowing of the spectrum with a FWHM of 4.6~kHz (see also
Fig.~\ref{noise_and_canonical_eit}).  Before the cell, we have
a Gaussian distribution of the spectral density
\begin{equation}
\label{spectr_in}
f_{\mathrm{in}}(\omega)=e^{-\frac{(\omega-\omega_0)^2}{\omega_w^2}}
\end{equation}
where $\omega_0$ is the average probe field frequency and
$\omega_w$ is the width of the spectral density spectrum.
However, after the cell, the probe field has a Lorentzian
distribution
\begin{equation}
\label{spectr_out}
f_{\mathrm{out}}=\frac{\gamma_{n}^2}{(\omega-\omega_0)^2+\gamma_{n}^2}
\end{equation}
where $\gamma_{n}$ is the width of the transmitted spectral
density spectrum.

	In Fig.~\ref{noise_and_canonical_eit} we compare
transmission spectra taken under two experimental conditions:
First, we measure the usual EIT resonance with coherent drive
and probe fields by scanning the two-photon detuning between the
two coherent fields.  Second, we use the phase-noise broadened
probe and measure the spectral density of the transmitted
light.  Note that the measured spectra after normalization
coincide within our experimental accuracy, which means that
$\gamma_n=\gamma_{EIT}$ where $\gamma_{EIT}$ is the width of
the EIT measured with the coherent probe and drive fields.

\begin{figure}
    \includegraphics[
        width=1.00\columnwidth
    ]
    {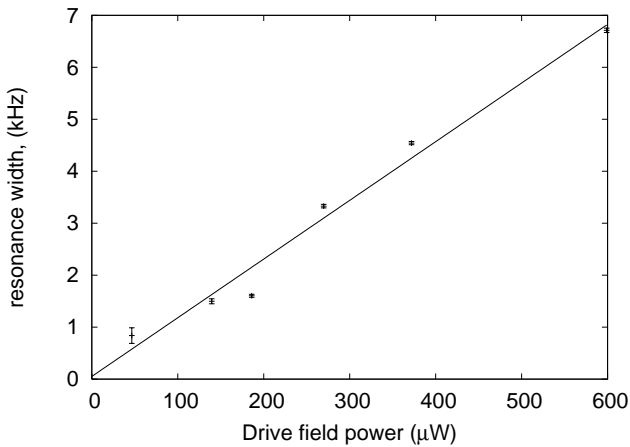}
    \caption{
	Dependence of spectral width of the transmitted probe
	field (see Fig.~\ref{noise_and_canonical_eit}) vs
	power of the drive laser under the condition of EIT resonance.
Thus, the coherence time of the probe field can be controlled by the power of
	drive field.
        \label{noise_width_vs_power}
    }
\end{figure}

	It is interesting to note that we significantly
increase the correlation time of the output probe field
$\tau_{\mathrm{out}}$ with respect to the correlation
time of the in-going probe field $\tau_{\mathrm{in}}$.  The
characteristic coherence time $\tau_{\mathrm{in}}=2/\omega_w$
for the input probe field is much smaller than the
characteristic coherence time of the output radiation
$\tau_{\mathrm{out}}=2/\gamma_n=2/\gamma_{EIT}$, as $\omega_w
\gg \gamma_{EIT}$.  Thus, we can create a source of radiation
with controllable coherence time.  This point is illustrated in
Fig.~\ref{noise_width_vs_power}, where we measure the dependence
of the transmitted spectral width of the signal as a function
of the power of the drive laser.

	In order to understand the surprising fact that the
spectral density of the beating signal does not significantly
depend upon the spectral widths of the drive and the probe
fields, we first recall that the phase-noise broadened probe
and drive field can be written as
$
E_p(t)=E_{p0} e^{-i( \omega_p t + \varphi_p(t) )},
E_d(t)=E_{d0} e^{-i( \omega_d t + \varphi_d(t) )}.
$
If the phase fluctuations are random, then
$\langle \varphi_d \rangle =0$,
$\langle \varphi_p \rangle =0$,
$\langle \dot{\varphi_d}(t) \dot{\varphi_d}(t') \rangle = 2 D_d \delta(t'-t)$,
and
$\langle \dot{\varphi_p}(t) \dot{\varphi_p}(t') \rangle = 2 D_p \delta(t'-t)$.
Thus, as been shown in~\cite{zubairy94pra},
the EIT resonance is broadened by the quantity $D_p+D_d$.

	Our experimental results can be explained by using
a simplified three-level structure for $^{87}$Rb atoms as depicted
in Fig.~\ref{level-structure}.  We let levels $a$, $b$, and $c$
correspond to $^{87}$Rb atomic levels $5P_{1/2}F=2$, $5S_{1/2}F=1$, and
$5S_{1/2}F=2$, respectively.  The driving field couples levels
$a$ and $c$ and the probe field couples levels $a$ and $b$.  The
interaction picture Hamiltonian for the system can be written as
\begin{equation}
\hat H = \hbar\Omega_d e^{i\Delta_d t}|a\rangle\langle c| +
\hbar\Omega_p e^{i\Delta_p t}|a\rangle\langle b|
+ h.c.
\end{equation}
where $\Delta_p = \omega_{ab} - \nu_p$ and $\Delta_d =
\omega_{ac} - \nu_d$ are the detunings of probe and drive
fields from the atomic transitions $a$-$b$ and $a$-$c$,
respectively, $\nu_p$ and $\nu_d$ are the frequencies of
the probe and drive fields, $\Omega_d=\wp_{ac}E_p/\hbar$
and $\Omega_p = \wp_{ab}E_p/\hbar$ are the Rabi frequencies,
and $\wp_{ab}$ and $\wp_{ac}$ are the dipole moments of the
$a$-$b$ and $a$-$c$ transitions.

	We are interested in finding the spectrum of the beat
signal $S(t,z)=\Omega_p(t,z)\Omega^*_d(t,z)$ that 
is governed by the propagation equation along $z$ axis ($0<z<L=2.5$ cm)
\begin{equation}
\label{}
\frac{\partial S} {\partial z}= -i \eta \rho_{ab} \Omega^*_d
+ i \eta \rho_{ca} \Omega_p,
\end{equation}
where $\rho_{ab}$ and $\rho_{ca}$ are the density matrix
elements; $\eta=3\lambda^2 N\g_r/8\pi$; $N$ is the atomic density; 
$\g_r$ is the radiative decay from level $a$ to level $b$;
$\lambda=2\pi c/\w_{ab}$.

The adiabatic approximation can be used to find the
coherences $\rho_{ab}$ and $\rho_{ca}$,	because the bandwidth of 
the phase-noise is less than relaxation rate of these coherences. 
Under these conditions,
the propagation of the correlation function for the beat signal,
$R(\tau,z)=\langle S(t,z) S(t+\tau,z) \rangle$, 
and time dependence of the correlation of beat signal and spin coherence 
$G(\tau,z)=\langle S(t,z) \rho_{cb}(t+\tau,z) \rangle$
whose space and time evolution is given by
\begin{eqnarray}
\label{eS}
\frac{\partial} {\partial z} R(\tau,z)= 
  2\eta \left(\frac{n_{ab}}{\Gamma_{ab}}-\frac{n_{ca}}{\Gamma_{ca}}\right)
  R(\tau,z) - \\\nonumber
  2\eta \left(\frac{|\Omega_d|^2}{\Gamma_{ab}}
              + \frac{|\Omega_p|^2}{\Gamma_{ca}}\right) G(\tau,z),
\end{eqnarray}
%
%
\begin{eqnarray}
\label{erho}
\frac{\partial}{\partial \tau}{ G(\tau,z)} =
%
\left(\frac{n_{ab}}{\Gamma_{ab}}-\frac{n_{ca}}{\Gamma_{ca}}\right)
	R(\tau,z) - && \\
\nonumber
\left(\Gamma_{cb} +
 \frac{|\Omega_d|^2}{\Gamma_{ab}} + \frac{|\Omega_p|^2}{\Gamma_{ca}}
  \right) G(\tau,z).
\end{eqnarray}
Here $\Gamma_{ab} = \gamma_{ab} + i\Delta_p$ and $\Gamma_{ca}
= \gamma_{ac} - i\Delta_{ac}$ where $\gamma_{ab}$
and $\gamma_{ac}$ are the relaxation rates of atomic
coherences $\rho_{ab}$ and $\rho_{ac}$ respectively.  Also
$n_{ab}=\rho_{aa}-\rho_{bb}$ and $n_{ca}=\rho_{cc}-\rho_{aa}$
are the population differences.  We have assumed a slow
variation of atomic populations such that they do not change
appreciably during propagation.  A detailed analysis justifying
this approximation will be presented elsewhere.  It is clear
that under these approximations, the correlation function of
the beat signal (and hence the corresponding spectral density)
is independent of the phase fluctuations of the the drive and
probe fields. 

Let us note the important role of atomic coherence in the discussed processes.
Before we proceed to solving a set of coupled Eqs.(\ref{eS},\ref{erho}), note
that the atomic coherence plays an important role in this processes, once
induced it gives rise a term $G(\tau, z)$. 
In the adiabatic limit, one can obtain transparency behavior governed by
%
\begin{eqnarray}
\frac{\partial} {\partial z} R(\tau,z)= 
  2\eta \ds{{\cal N}
{
	\Gamma_{cb} R(\tau,z)
\over
\tilde\Gamma_{cb}
}},
\end{eqnarray}
if 
condition $|\Omega_d|^2\gg \Gamma_{ab}\Gamma_{cb}$ is met. 
Here we introduce 
$\tilde\Gamma_{cb} = \Gamma_{cb} +
\frac{|\Omega_d|^2}{\Gamma_{ab}} +
\frac{|\Omega_p|^2}{\Gamma_{ca}}$ and 
${\cal N} = 
\frac{n_{ab}}{\Gamma_{ab}}-\frac{n_{ca}}{\Gamma_{ca}}$.

	In order to determine the spectral density of the beat
signal $I_{\omega}$, we recall the definitions
$$
\label{eq4}
R(\tau,z)
= \int{I_{\omega}(z) e^{-i \omega \tau } d \omega }, \;\;
%
%
G(\tau,z)=
\int{\rho_{\omega}(z) e^{-i \omega \tau } d \omega }.
$$
Substituting these expressions for 
$R(\tau,z)
$ and
$
G(\tau,z)$
into Eqs.~(\ref{eS}) and (\ref{erho}) we find
\begin{equation}
\label{eq6}
\rho_{\omega}=\frac{
{\cal N}
I_{\omega} }
{\tilde\Gamma_{cb} -i\omega
}
%
\mbox{ and }
%
\frac{\partial} {\partial z} I_{\omega}=2 \eta \frac{
    \left(\Gamma_{cb} - i \omega \right)
    {\cal N}
}
{\tilde\Gamma_{cb} -i\omega
}
I_{\omega} ~.
\end{equation}

	For the simplest case, with a weak probe and a strong
drive, $|\W_d|\gg |\W_p|$ and all population remains in state
$b$, so that $n_b=1$, $n_a=n_c=0$.  Taking Doppler broadening
into account by integration over velocity distribution leads
to changing the homogeneous width $\g$ ($2\cdot 10^7$ s$^{-1}$) 
to the Doppler width $\Delta_W$ ($2\pi 500$ MHz).
%
%
Also, there is narrowing because the medium is optically 
thick~\cite{lukin97prl}.
The spectral density of the beat signal is then given by
\begin{equation}
I_\w(z) =
 I_\w(0) \exp\left[
   \frac{-\eta z\w^2
    }{
     \Delta_W\left(
       \left( \frac{|\W|^2}{\Delta_W}\right)^2 + \w^2
         \right)
      }\right]~.
\end{equation}
with the following expression for the spectral width of
beating signal:
\begin{equation}
\Delta\w_{bs} = \ds{|\W|^2\over\Delta_W\sqrt{\ds{\eta
z\over\Delta_W} -1}} ~.
\end{equation}
These expressions agree very well with our experimental results, namely,
it gives a linear dependence of the spectral width on the driving power, and
the same slope for the experimental parameters and a density of atoms of the
order of $3\; 10^{11}$ cm$^{-3}$ which
corresponds to the cell temperature. 
Note, for the case of monochromatic drive and broadened probe field, 
the spectral width of the beating signal coincides with the spectral width of 
the EIT for a monochromatic probe field. 
The results obtained here can be interpreted in some sense that the 
probe field passing the cell and the atomic coherence in the cell are 
strongly correlated (see Eq.(8)) \cite{polzik}. 
It is also worth to mention that there are interesting aspects of the problem
was studied recently in~ where it has been demonstrated
that, when probe and drive intensities are the same order of magnitude,
the noise of probe and drive fields are strongly 
correlated~\cite{sautenkov05pra}, and 
phase noise of one laser is transfered to the second laser~\cite{huss04prl}. 
Also we would like to underline that the results obtained here go beyond
classical EIT treatment~\cite{zubairy94pra} where the broadening of EIT
resonance was predicted. 

	In conclusion, we have experimentally observed that
the width of the beat signal between a coherent drive and
broad-band probe field is greatly reduced (more than 1000 times)
by propagation through a cell containing $^{87}$Rb vapor.  We analyzed
the modification of a broad emission spectral line after
transmission through a coherently prepared resonant medium.
The final spectral lineshape is defined by the spectral shape
of EIT resonance. The applications of the obtained results 
can be light sources (including a single photon sources)~\cite{lukin05} 
with controllable coherent time of radiation, 
and notch filters with 
subnaturally narrow transmission band that are in great demand for 
many practical applications to background suppression in imaging
including astrophysics and environmental imaging~\cite{filters-miles}. 


	We thank S.\ E.\ Harris, M.D. Eisaman, M.D.\ Lukin, P.\ Meystre, and 
A.S.\ Zibrov for useful discussions and gratefully acknowledge 
the support from the Office of Naval Research, 
the Air Force Research Laboratory
(Rome, NY), Defense Advanced Research Projects Agency, and 
the Robert A.\ Welch Foundation (Grant No. A1261).

\end{document}